# On comparing sums of square roots of small integers [*]

Qi Cheng[†]

**Abstract**

Let $k$ and $n$ be positive integers, $n > k$. Define $r(n, k)$ to be the minimum positive value of
$$|\sqrt{a_1} + \cdots + \sqrt{a_k} - \sqrt{b_1} - \cdots - \sqrt{b_k}|$$
where $a_1, a_2, \cdots, a_k, b_1, b_2, \cdots, b_k$ are positive integers no larger than $n$. It is an important problem in computational geometry to determine a good upper bound of $-\log r(n, k)$. In this paper we prove an upper bound of $2^{O(n/\log n)} \log n$, which is better than the best known result $O(2^{2k} \log n)$ whenever $n \leq ck \log k$ for some constant $c$. In particular, our result implies a *subexponential* algorithm to compare two sums of square roots of integers of size $o(k \log k)$.

## 1 Introduction

In computational geometry, one often needs to compare lengths of two polygonal paths, whose nodes are on a integral lattice, and edges are measured according to the $L_2$ norm. The problem can be reduced to the problem of comparing two sums of square roots of integers. Most work in computational geometry assumes a model of real-number machines, where one memory cell can store one real number. It is assumed that an algebraic operation, taking a square root as well as a comparison between real numbers can be done in one operation. There is a straight-forward way to compare sums of square roots in real-number machines. But this model is not realistic, as shown in [6, 5].

If we consider the problem in the model of Turing machine, then we need to design an algorithm to compare two sums of square roots of integers with low bit complexity. One approach would be approximating the sums by decimal numbers up to a certain precision, and then hopefully we can learn which one is larger. Formally define $r(n, k)$ to be be the minimum positive value of
$$|\sqrt{a_1} + \cdots + \sqrt{a_k} - \sqrt{b_1} - \cdots - \sqrt{b_k}|$$

---

[*]This research is partially supported by NSF Career Award CCR-0237845.

[†]School of Computer Science, the University of Oklahoma, Norman, OK 73019, USA. Email: qcheng@cs.ou.edu.

where $a_1, a_2, \cdots, a_k, b_1, b_2, \cdots, b_k$ are positive integers no larger than $n$. The time complexity of the approximation approach depends directly on $-\log r(n, k)$, since an approximation of a sum of square roots of integers can be computed in time polynomial in the number of digits in the approximation. One would like to know if $-\log r(n, k)$ is bounded from above by a polynomial function in $k$ and $\log n$. If so, the approximate approach to compare two sums of square root of integers runs in polynomial time. Note that even if the lower bound of $-\log r(n, k)$ is exponential, it does not necessarily rule out a polynomial time algorithm.

Although this problem was put forward during the 1980s [3], progress has been scarce. In [1], it is proved that
$$-\log r(n, k) = O(2^{2k} \log n)$$
using the root separation method. This immediately gives us a polynomial time algorithm of comparing sum of square roots if $k$ is fixed. Qian and Wang [4] gave a constructive upper bound of $r(n, k)$ at $O(n^{-2k+\frac{3}{2}})$, which corresponds to a lower bound
$$-\log r(n, k) = \Omega(k \log n).$$
They conjecture that $-\log r(n, k) = \Theta(n^{\frac{1}{2} - 2^{k-2}})$.

There is a wide gap between the known upper bound and lower bound of $-\log r(n, k)$. Until the fundamental problem has been resolved, we can not even put the presumably easy problem such as Euclidean Minimum Spanning Tree problem in P, and the Euclidean Traveling Salesman problem in NP.

## 1.1 Our contribution

From the known upper bound of $-\log r(n, k)$, we conclude that there is a polynomial time algorithm to compare sum of square roots if $k$ is fixed. In this note, we consider the case in the other end of the spectrum when $k$ grows (almost) linear with $n$.

**Definition 1** *An integer $n$ is called square free, if there does not exist a prime $p$ such that $p^2$ divides $n$.*

It is well known that there are about
$$\frac{6n}{\pi^2} + O(\sqrt{n})$$
many square free integers less than $n$. If $a_1, a_2, \cdots, a_k, b_1, b_2, \cdots, b_k$ are distinct square free integers, then their square roots are linearly independent over the field of rational number $\mathbf{Q}$. So it is possible that $k$ and $n$ are linearly related. This case is also practically interesting. We often need to compare paths whose nodes are on an $l \times l$ integral grid. The distance between the lattice points are square roots of integers of size $O(l^2)$. There are $l^2$ many nodes in the grid, and if we select a dense subset out of the grid points, we arrive in the situation where $n$ is linear in $k$.

We obtain a lower bound of difference of two sums of square roots. Our lower bound beats the root separation bound as long as $n \leq ck \log k$ for some constant $c$. The corresponding upper bound on $-\log r(n, k)$ becomes *subexponential* when $n = o(k \log k)$, or more generally, if the square free parts of the numbers grow at rate $o(k \log k)$. Our bound implies a subexpontial algorithm, i.e., an algorithm with time complexity $2^{o(k)} \log n$, to compare two sums of square roots of small integers. The proof is also simple.

We begin the presentation of our result by defining the notion of multiplicative generators.

**Definition 2** *Given two set of positive integers $A$ and $B$, we say that $B$ multiplicatively generates $A$ if any number in $A$ can be written as a product of numbers from $B$ with repetition allowed.*

It is easy to see that $A$ multiplicatively generates itself, but for many sets, there exist much smaller sets which multiplicatively generate them. For example, all the square free number less than $n$ are generated by the set of primes less than $n$, whose cardinality is $O(n/\log n)$.

**Theorem 1** *(Main) Let $c_1, c_2, \cdots, c_k, d_1, d_2, \cdots, d_k$ be positive integers. Let*

$$A = \{a_1, a_2, \cdots, a_k, b_1, b_2, \cdots, b_k\}$$

*be the set of $2k$ positive square free integers. Assume that $c_i^2 a_i \leq n$ for all $1 \leq i \leq k$ and $d_i^2 b_i \leq n$ for all $1 \leq i \leq k$. Let $B$ be a set which multiplicatively generates $A$. Then*

$$|c_1\sqrt{a_1} + \cdots + c_k\sqrt{a_k} - d_1\sqrt{b_1} - \cdots - d_k\sqrt{b_k}| > (2k\sqrt{n})^{-2^{|B|}+1}.$$

Since $A$ generates itself, so this result recovers the best known lower bound on $r(n, k)$. In many cases, this result improves that bound, since $|B|$ can be smaller than $|A| = 2k$. It is possible that the cardinality of $B$ can be as small as $O(\log k)$, in which case, there is a polynomial time algorithm comparing two sums of square roots.

Our result shows that the multiplicative structure of $A$ affects the minimum possible value of $|c_1\sqrt{a_1} + c_2\sqrt{a_2} + \cdots + c_k\sqrt{a_k} - d_1\sqrt{b_1} - d_2\sqrt{b_2} - \cdots - d_k\sqrt{b_k}|$, which appears to be unknown before. In particular, we show that the root separation lower bound $2^{O(k)} \log n$ of $-\log r(n, k)$ is *not* tight, at least, when $n$ is linear in $k$. It is still possible that when $n$ is much larger than $k$, the root separation bound becomes tight. Our result indicates that to achieve the root separation bound, it is important to select the numbers $a_1, a_2, \cdots, a_k, b_1, b_2, \cdots, b_k$ such that they are pairwise relatively prime.

## 2 The proof

Let $F = \mathbf{Q}(x_1, x_2, \cdots, x_m)$ be the function field over $\mathbf{Q}$ with indeterminate $x_1, x_2, \cdots, x_m$. Consider a field extension $K = \mathbf{F}[y_1, y_2, \cdots, y_m]/(y_1^2 - x_1, \cdots, y_m^2 - x_m)$ of $F$. It is a linear space of dimension $2^m$ over $\mathbf{Q}(x_1, x_2, \cdots, x_m)$, one of whose bases is

$$\{B_S = \prod_{i \in S} y_i | S \subseteq \{1, 2, \cdots, m\}\}.$$

The Galois group $G$ of $K$ over $F$ has order $2^m$. For any subset $S$ of $\{1, 2, \cdots, m\}$, define $\sigma_S \in G$ recursively as follows:

1. If $S = \emptyset$, $\sigma_S$ is the identity element.

2. If $|S| = 1$, then
$$\sigma_{\{i\}}(y_j) = \begin{cases} -y_j & \text{if } i = j \\ y_j & \text{if } i \neq j \end{cases}$$

3. If $|S| > 1$, $\sigma_S = \prod_{i \in S} \sigma_{\{i\}}$.

We have $\sigma_{S'}(B_S) = (-1)^{|S' \cap S|} B_S$ and $G = \{\sigma_S | S \subseteq \{1, \cdots m\}\}$

**Lemma 1** *Let $\{\alpha_S | S \subseteq \{1, 2, 3, \cdots, m\}\}$ be a set of $2^m$ integers. The norm of $\sum_{S \subseteq \{1,2,\cdots,m\}} \alpha_S B_S$, denoted by*
$$N_{K/F}(\sum_{S \subseteq \{1,2,\cdots,m\}} \alpha_S B_S)$$
*is a polynomial in $\mathbf{Z}[x_1, x_2, \cdots, x_m]$.*

*Proof:* By definition,
$$N_{K/F}(\sum_{S \subseteq \{1,2,\cdots,m\}} \alpha_i B_i) = \prod_{\sigma \in G} \sigma(\sum_{S \subseteq \{1,2,\cdots,m\}} \alpha_i B_i) \tag{1}$$
$$= \prod_{S' \subseteq \{1,2,\cdots,m\}} (\sum_{S \subseteq \{1,2,\cdots,m\}} \alpha_S \sigma_{S'}(B_S)) \tag{2}$$
$$= \prod_{S' \subseteq \{1,2,\cdots,m\}} (\sum_{S \subseteq \{1,2,\cdots,m\}} (-1)^{|S \cap S'|} \alpha_S B_S). \tag{3}$$

The norm must be an element in $F = \mathbf{Q}(x_1, x_2, \cdots, x_m)$. On the other hand, if we expand the product in the right hand side, it reduces to $\sum_{S \subseteq \{1,2,\cdots,m\}} \beta_S B_S$, where $\beta_S \in \mathbf{Z}[x_1, x_2, \cdots, x_m]$ for any $S \subseteq \{1, 2, \cdots, m\}$. Hence we must have $\beta_S = 0$ for $|S| \geq 1$. Thus
$$N_{K/F}(\sum_{S \subseteq \{1,2,\cdots,m\}} \alpha_S B_S) = \beta_\emptyset,$$
which is a polynomial in $\mathbf{Z}[x_1, x_2, \cdots, x_m]$. $\square$

Define the polynomial
$$f_{\alpha_\emptyset, \alpha_{\{1\}}, \alpha_{\{2\}}, \cdots, \alpha_{\{1,2,\cdots,m\}}}(x_1, x_2, \cdots, x_m) = N_{K/F}(\sum_{S \subseteq \{1,2,\cdots,m\}} \alpha_S B_S) \in \mathbf{Z}[x_1, x_2, \cdots, x_m].$$

Now we are ready to prove the main theorem.

*Proof:* (of the main theorem) Denote $|B|$ by $m$. Assume that $B = \{h_1, h_2, \cdots, h_m\}$. There is a natural ring homomorphism
$$\psi : \mathbf{Q}[x_1, x_2, \cdots, x_m, y_1, y_2, \cdots, y_m] \to \mathbf{Q}(\sqrt{h_1}, \sqrt{h_2}, \cdots, \sqrt{h_m})$$

by letting $\psi(y_i) = \sqrt{h_i}$ and $\psi(x_i) = h_i$ for $1 \le i \le m$.

Fix an order among all the subsets of $\{1, 2, \cdots, m\}$. Define $B'_S = \prod_{i \in S} h_i$, and define $\alpha_S$ as

$$\alpha_S = \begin{cases} c_j & \text{if } S \text{ is the first set such that } a_j = B'_S \\ -d_j & \text{if } S \text{ is the first set such that } b_j = B'_S \\ 0 & \text{Otherwise} \end{cases}$$

We have

$$c_1\sqrt{a_1} + \cdots + c_k\sqrt{a_k} - d_1\sqrt{b_1} - \cdots - d_k\sqrt{b_k} = \sum_{S \subseteq \{1,2,\cdots,m\}} \alpha_S B'_S$$

and

$$f_{\alpha_\emptyset, \cdots, \alpha_{\{1,2,\cdots,m\}}}(h_1, h_2, \cdots, h_m) = \prod_{S' \subseteq \{1,2,\cdots,m\}} (\sum_{S \subseteq \{1,2,\cdots,m\}} (-1)^{|S' \cap S|} \alpha_S B'_S)$$

because of the ring homomorphism. The integer

$$f_{\alpha_\emptyset, \cdots, \alpha_{\{1,2,\cdots,m\}}}(h_1, h_2, \cdots, h_m) \ne 0$$

since $\sqrt{a_1}, \sqrt{a_2}, \cdots, \sqrt{a_k}, \sqrt{b_1}, \sqrt{b_2}, \cdots, \sqrt{b_k}$ are linear independent over $\mathbf{Q}$. So

$$|\prod_{S' \subseteq \{1,2,\cdots,m\}} (\sum_{S \subseteq \{1,2,\cdots,m\}} (-1)^{|S' \cap S|} \alpha_S B'_S)| \ge 1$$

Thus

$$|c_1\sqrt{a_1} + \cdots + c_k\sqrt{a_k} - d_1\sqrt{b_1} - \cdots - d_k\sqrt{b_k}| \qquad (4)$$

$$\ge \frac{1}{\prod_{|S'| \ne \emptyset} (\sum_{S \subseteq \{1,2,\cdots,m\}} (-1)^{|S' \cap S|} \alpha_S B'_S)} \qquad (5)$$

$$\ge \frac{1}{(2k\sqrt{n})^{2^{|B|}-1}}. \qquad (6)$$

□

The proof relies on the fact that the norm is a nonzero integer, thus has absolute value greater than 1. Every factor in the definition of the norm is not too large (less than $2k\sqrt{n}$ in our case), so the smallest factor should not be too small. The technique has been used in several papers, for example, see [2]. The estimation depends primarily on the number of factors in the definition of the norm.

## 3   A corollary from the main theorem

**Theorem 2** *Let $c_1, c_2, \cdots, c_k, d_1, d_2, \cdots, d_k$ be positive integers. Let $a_1, a_2, \cdots, a_k, b_1, b_2, \cdots, b_k$ be distinct square free positive integers less than $m$. Assume that $c_i^2 a_i \le n$ for all $1 \le i \le k$ and $d_i^2 b_i \le n$ for all $1 \le i \le k$. Then*

$$|c_1\sqrt{a_1} + \cdots + c_k\sqrt{a_k} - d_1\sqrt{b_1} - \cdots - d_k\sqrt{b_k}| > (2k\sqrt{n})^{-2^{O(m/\log m)}}$$

*Proof:* It is well known that the number of primes less than $m$ is $O(m/\log m)$. The set of primes less than $m$ generates all the positive integers less than $m$. The theorem follows from the main theorem. □

**Corollary 1** $-\log r(n,k) = 2^{O(n/\log n)} \log n$

## 4 Conclusion remarks

In this paper, we prove an upper bound of $2^{O(n/\log n)} \log n$ for $-\log r(n,k)$, by exploring the fact that the algebraic degree of sum of $2k$ square free positive integers can be much less than $2^{2k}$. We suspect that $2^{O(k/\log k)} \log n$ type of upper bound holds for much large $n$, and leave it as an open problem.